\documentclass[%
twocolumn,
 reprint,
superscriptaddress,
 amsmath,amssymb,
 aps,
prd,
longbibliography,
]{revtex4-1}

\usepackage{aas_macros}
\usepackage{graphicx}% Include figure files
\usepackage{dcolumn}% Align table columns on decimal point
\usepackage{bm}% bold math
\usepackage{hyperref}% add hypertext capabilities
\usepackage{orcidlink}

\usepackage{color}
\usepackage[mathlines]{lineno}% Enable numbering of text and display math
\usepackage{tabularx}
\usepackage{graphicx}
\usepackage{subfigure}
\usepackage[normalem]{ulem}
\usepackage[export]{adjustbox} 
\usepackage{url}

\hypersetup{
    pdfnewwindow=true,    % links in new window
    colorlinks=true,      % false: boxed links; true: colored links
    linkcolor=blue,       % color of internal links
    citecolor=blue,       % color of links to bibliography
    filecolor=blue,       % color of file links
    urlcolor=blue         % color of external links
}

\usepackage{natbib}

\begin{document}
\title{Improving HAWC dark matter constraints with Inverse-Compton Emission}
\author{Dylan M. H. Leung, \orcidlink{0009-0007-5539-2151}}
\email{leungmanhei@link.cuhk.edu.hk}
\thanks{\scriptsize \!\!\href{https://orcid.org/0009-0007-5539-2151}{orcid.org/0009-0007-5539-2151}}
\affiliation{Department of Physics, The Chinese University of Hong Kong, Shatin, New Territories, Hong Kong, China}

\author{Kenny C. Y. Ng, \orcidlink{0000-0001-8016-2170}}
\email{kcyng@cuhk.edu.hk}
\thanks{\scriptsize \!\! \href{http://orcid.org/0000-0001-8016-2170}{orcid.org/0000-0001-8016-2170}}
\affiliation{Department of Physics, The Chinese University of Hong Kong, Shatin, New Territories, Hong Kong, China}

\date{December 14, 2023}

\begin{abstract}

The particle nature of dark matter (DM) has been a long-lasting mystery. Many models suggest that DM could decay or self annihilate into standard model particles, and thus could be a source of gamma rays in the sky. 
The High Altitude Water Cherenkov (HAWC) observatory has yielded some of the strongest limits in searches of DM decay or annihilation. Building on the flux limits provided by the HAWC collaboration in 2018, we consider the effects of additional components from Galactic secondary Inverse-Compton scatterings and extragalactic DM distributions. We find that these effects can significantly improve the DM constraints, up to an order of magnitude in some cases. This highlights the importance of considering secondary contributions in detail in LHAASO-WCDA and SWGO in the future.

\end{abstract}

\maketitle

\section{Introduction} \label{sec:Introduction}
\par
The evidence of dark matter (DM) has been well established through its significant gravitational effects in many observations over the last few decades. Meanwhile, numerous ongoing efforts are dedicated to searching for its non-gravitational interactions and examining its possible candidates~\cite{Ellis:1983ew, Bertone:2004pz}. Significant progress has been made in constraining various potential DM properties, e.g., the decay lifetime of Very Heavy dark matter \cite{Kalashev:2016cre, Chianese:2021jke, Ishiwata:2019aet} and the annihilation cross section for Weakly Interactive Massive Particles~\cite{Schumann:2019eaa}.
\par

If DM can decay or annihilate to standard model final states within the age of the current Universe ~\cite{Ibarra:2007wg, Kolb:1998ki, Halverson:2016nfq, Buch:2015iya, Bauer:2020jay}, stable final-state products, such as gamma rays or neutrinos, should leave some imprint in the sky. 
Such signals can be searched using gamma-ray or neutrino telescopes. In this case, the signal is expected to be highest at sky regions with high concentration of DM content, such as near the center of the galactic halo. 

There have been many attempts at searching for the high-energy gamma-ray or neutrino signals around the ${\rm GeV}-{\rm TeV}$ energy band. These include HAWC~\cite{HAWC:2014ycj} searching for signals from the inner galaxy~\cite{HAWC:2017udy}, HESS, MAGIC, Fermi-LAT, HAWC, and VERITAS searching for signals from satellite galaxies~\cite{HESS:2014zqa, MAGIC:2016xys, VERITAS:2017tif, HAWC:2017mfa}, IceCube diffuse neutrino searches~\cite{Chianese:2019kyl, IceCube:2023ies}, Fermi-LAT diffuse gamma-ray and inner galaxy search, LHAASO galactic halo search~\cite{LHAASO:2022yxw}, and more.  No conclusive detection has been made, and thus limits on the DM properties are obtained from these works.

Previous works by the HAWC Collaboration~\cite{HAWC:2017udy} obtained constraints on DM lifetime and annihilation cross sections based on observations toward the inner galaxy~\cite{Abeysekara:2017wzt}. However, only prompt emission is considered.  The goal of this work is to re-examine the lifetime and cross-section limits by including secondary emissions from Inverse Compton (IC) scattering with the interstellar radiation field, as well as the extragalactic prompt and secondary components. 

We review the formalism to calculate the prompt and secondary gamma-ray flux in section~\ref{sec:Formulation}. In Section~\ref{sec:Method}, we provide a brief overview of the observation method and working principle of the HAWC telescope, as well as an outline of the statistical analysis used to derive the final lifetime and cross-section limits. In Section~\ref{sec:result}, we discuss the improvements made in this study and compare the obtained limits with those from other studies. Finally, in Section~\ref{sec:Conclusions}, we conclude our current progress and outline possible future works.

\section{Gamma Rays from DM Decay and Annihilation} 
\label{sec:Formulation}

To obtain the expected gamma-ray flux from DM decay and annihilation, we consider prompt and secondary emission from Galactic and extragalactic DM distributions. 

\begin{figure}[t!]
    \centering
    \includegraphics[width=83mm]{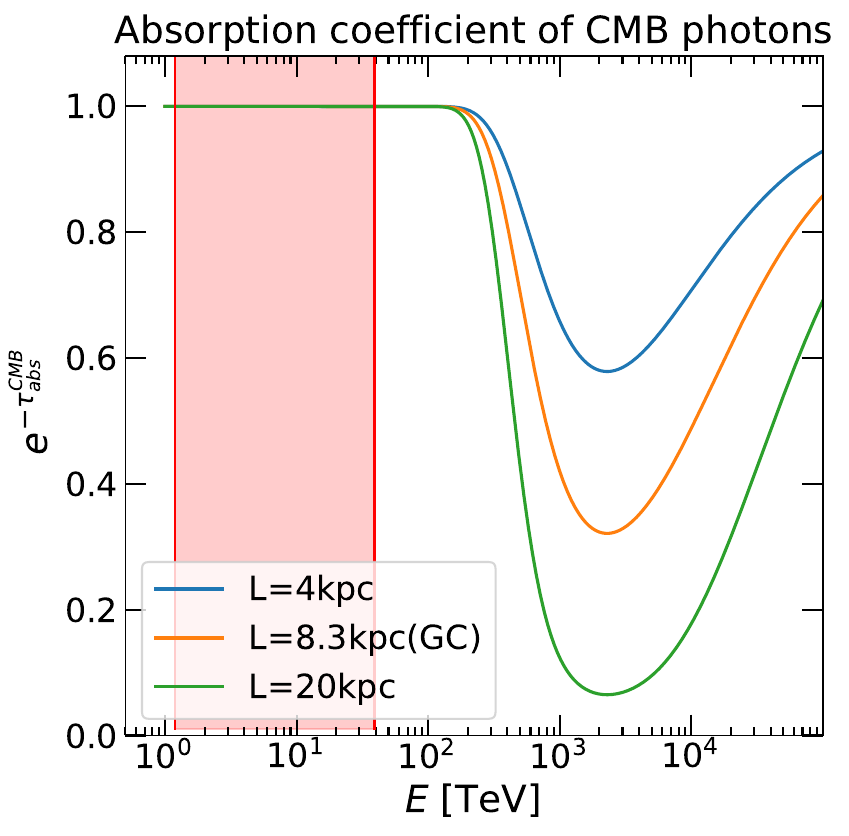}
    \caption{The absorption term of gamma rays with CMB photon plotted versus gamma-ray energy.  $L$ here indicates the traveling distances of 4kpc, 8.3kpc~(GC distance from the Earth), and 20kpc. The red shaded region is the energy range considered in this work.
    }
    \label{fig:.abs}
\end{figure}

\subsection{Galactic Prompt Component}
The prompt flux of DM decay or annihilation comes from the gamma rays directly produced by the final-state particles. E.g., for $DM\rightarrow b\bar{b}$ channel, the further hadronization and particle decays from the quarks would lead to the production of gamma rays. For channels like $DM\rightarrow e^{+}e^{-}$, gamma rays can still be produced promptly by high-order corrections to the process. 

To obtain the gamma-ray emission, it is also necessary to specify the distribution of DM in the galaxy. In this study, we adopt the popular Navarro-Frenk-White (NFW) profile~\cite{Navarro:1996gj} for the galactic DM profile
\begin{equation}
    \rho_{DM}(r) = \frac{\rho_{s}}{(r/r_{s})(1+r/r_s)^{2}}\,,
\end{equation}
where the scale radius is $r_{s} =20\,{\rm kpc}$ \cite{Nesti:2013uwa}, the scale density is $\rho_s=0.318\,{\rm GeV cm^{-3}}$~\cite{HAWC:2017udy}, which correspond to the local DM density to be about $0.38\,{\rm GeV cm^{-3}}$. The distance $r$ from the galactic center (GC) can be calculated as a function of $x$ (line-of-sight distance from Earth) and $\theta$ (angle between the line-of-sight and the GC) as
\begin{equation}
    r(\theta, x) = \sqrt{R_{sun}^{2}-2 x R_{sun}\cos(\theta)+x^{2}}\,,
\end{equation}
where $R_{sun} = 8.3\,{\rm kpc}$ is the distance between the Sun and the GC.

There are also other possible DM profiles, such as the Einasto profile~\cite{Navarro:2008kc} and Burkert profile~\cite{Burkert:1995yz}. However, for the distance range considered in this study ($\rm 1kpc < r < 10kpc$), all three profiles generally agree well over this region. Therefore, the choice of profile is less relevant to the calculation of the gamma-ray flux in this study~\cite{HAWC:2017udy}. 

The prompt gamma-ray flux from DM decay and annihilation is then
\begin{eqnarray}\label{eq:PFlux}
\nonumber    
\frac{d^{2}\phi_{prompt}}{dEd\Omega}\Big|_{decay} &=& D\frac{1}{4\pi M_{DM}\tau}\frac{dN_\gamma}{dE}e^{-\tau_{abs}^{ISRF}}\,, \\
    \frac{d^{2}\phi_{prompt}}{dEd\Omega}\Big|_{ann} &=& J\frac{\langle\sigma v\rangle}{8\pi M_{DM}^2}\frac{dN_\gamma}{dE}e^{-\tau_{abs}^{ISRF}}\,, 
\end{eqnarray}
where $M_{DM}$ is the DM mass, $\tau$ is the decay lifetime, $\langle\sigma v\rangle$ is the annihilation cross section, $dN_{\gamma}/dE$ is the gamma-ray spectrum per decay/annihilation. (Note that decay and annihilation spectrum differs by a factor of 2 in central of mass energy.)  The D-factor ($D$) and J-factor ($J$) represents the line-of-sight integral of the DM content at a certain observational angle, 
\begin{eqnarray}\label{eq:dfactor}
\nonumber 
    D=\int_{l.o.s} d x \rho_{DM}(r(\theta, x))\,,  \\
    J=\int_{l.o.s} d x \rho_{DM}^2(r(\theta, x))\, .
\end{eqnarray}
For $dN_{\gamma}/dE$, we use the HDM package~\cite{Bauer:2020jay}, which takes into account electroweak corrections that become increasingly important at high energies. 

The additional exponential term  $e^{-\tau_{abs}^{ISRF}}$ represents the absorption of gamma rays as they propagate through interstellar space, taking into account the pair production process $\gamma\gamma \rightarrow e^{+}e^{-}$ with background photons. The explicit form of the absorption term can be found in Ref.~\cite{Esmaili:2015xpa}. Fig.~\ref{fig:.abs} illustrates the absorption term under the Cosmic Microwave Background~(CMB) as a function of photon energy, gamma ray sources at various distances from the Earth are shown.  
Notably, the absorption becomes most significant around the energy level of $10^{3}\, {\rm TeV}$. However, in this study, the photon energy range of interest is within the region bounded by the red region $[10^3-4\times10^4]\,{\rm GeV}$. As the absorption term consistently remains close to a value of 1, it is safely ignored in this work.
Similar arguments can be made for the absorption terms under star light (SL) and Infra Red (IR) radiation, where the absorption coefficient remains fairly close to 1 within our energy range of interest.~\cite{Esmaili:2015xpa}.

\begin{figure}[t!]
    \centering
    \includegraphics[width=88mm]{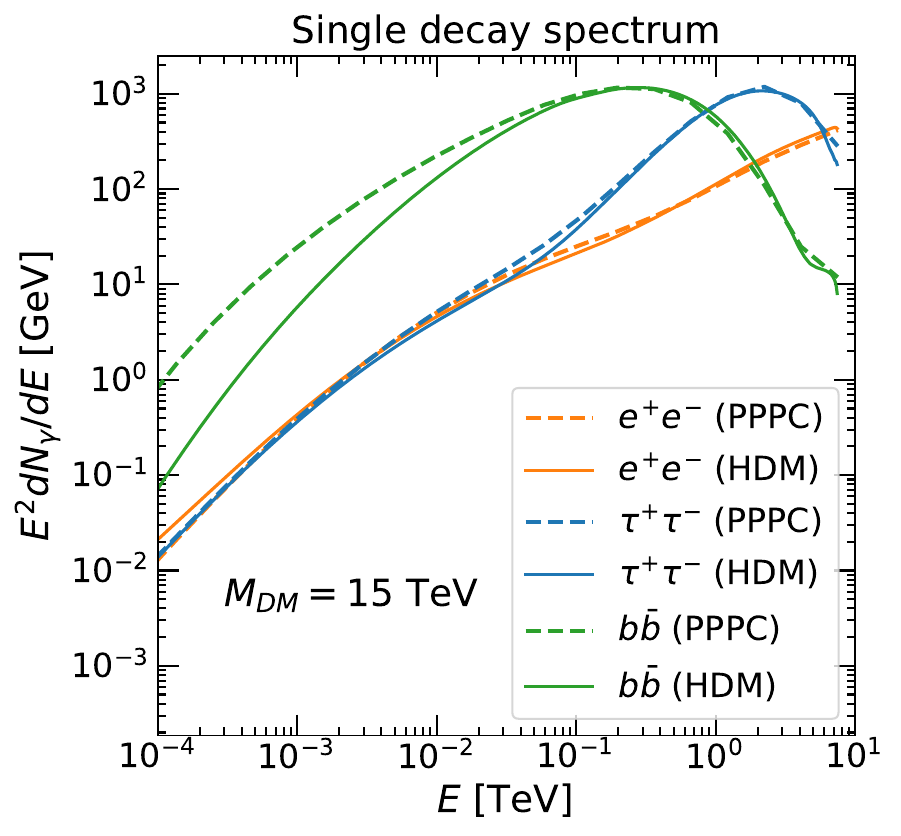}
    \caption{The single decay spectra from PPPC4DMID \cite{Buch:2015iya} and HDM \cite{Bauer:2020jay} for three channels with DM mass set to 15\,TeV. The HDM results are adopted in this work. The two codes generally agrees well with each other. }
    \label{fig:hdm}
\end{figure}

Fig.~\ref{fig:hdm} shows the single decay spectrum ($dN_{\gamma}/dE$) obtained from HDM for the several channels. 
For comparison, we also compare the HDM results with PPPC4DMID, another popular code for computing the spectrum~\cite{Cirelli:2010xx}. The two codes generally agrees well with each other, especially at the important part at high energies and at the peak of the spectrum. 

\subsection{Galactic Secondary Component: IC emission}
In addition to the prompt flux, electrons (and positrons) produced by DM decay/annihilation could also upscatter background photons into gamma-ray energies. These photon background include the CMB, IR, and SL. In this study, we adopt the full interstellar radiation field~(ISRF), $n_\gamma^{ISRF} = n_\gamma^{CMB} + n_\gamma^{IR} + n_\gamma^{SL}$.

The IC flux depends on the distribution of electrons produced by DM decay/annihilation. To obtain that, it is necessary to consider the effect of electron propagation in the Galactic environment. This is often done by considering the diffusion-loss equation
\begin{equation}\label{eq:DLE}
    \frac{\partial n_e}{\partial t} - \nabla (I(E,x) \nabla n_e) - \frac{\partial}{\partial E}(b(E) n_e) = q(E)\,,
\end{equation}
where $n_e(r,E_e)$ is the electron number density per unit energy, $I$ is the diffusion coefficient, $b(E)$ is the continuous energy loss rate, and $q(E)$ is the electron source term.

To solve Eq.~\ref{eq:DLE}, we adopt the same assumption made in Ref.~\cite{Cirelli:2009vg}, where the diffusion term is neglected and assuming time equilibrium ($\frac{\partial n_e}{\partial t} = 0$). This is justified by considering that the energy loss term rapidly grow with energy ($b\propto E^{2}$), while the diffusion coefficient decrease with energy to some small power.  
Therefore, in the very-high-energy regime, electrons can only travel a small finite range of distance within the characteristic time due to radiative losses.

The solution for the diffusion-loss equation can then be easily obtained,
\begin{equation}\label{eq:SDLE}
    n_{e}(r, E_e)=\frac{1}{b(E_e)}\int^{M_{DM}}_{E_e}q(E')dE'\,,
\end{equation}
where the source term $q(r, E)$ from DM decay/annihilation is
\begin{eqnarray}
\nonumber
    q_{decay}(r, E)_=\frac{\rho_{DM}(r)}{M_{DM}\tau}\frac{dN_e}{dE}\,, \\
    q_{ann}(r, E)=\langle\sigma v\rangle\frac{\rho_{DM}^2(r)}{2M_{DM}^2}\frac{dN_e}{dE}\,,
\end{eqnarray}
and $\frac{dN_e}{dE}$ is the electron spectrum per decay/annihilation, which we also obtain from HDM. 

In Eq.~\ref{eq:SDLE}, the term $b(E)$  represents the total energy loss coefficient, which includes the IC scattering and the synchrotron energy loss in the presence of the galactic magnetic field:
\begin{equation}
    b(E_e) = -\frac{dE_e}{dt} = b_{IC}(r, E_e)+b_{SYN}(\rho, z, E_e)\,.
\end{equation}
The energy loss due to IC scattering for a given electron energy, $E_e$, is~\cite{Buch:2015iya}
\begin{multline}
    b_{IC}(r, E_e) = 3\sigma_{T} \int^{\infty}_{0}E_{\gamma} dE_{\gamma} \int^1_{\frac{1}{4\gamma^2}}dq n_{\gamma}(r,E_{\gamma})\\
    \frac{(4\gamma^{2}-\Gamma_{\epsilon})-1}{(1+\Gamma_{\epsilon}q)^3}
    \left[2q\ln{q} +q+1
    -2q^{2}+\frac{(\Gamma_{\epsilon} q)^{2}(1-q)}{2(1+\Gamma_{\epsilon}q)} \right],
\end{multline}
where $\sigma_T$ is the classical Thomson cross-section, $\Gamma_{\epsilon}=4 E_{\gamma} \gamma/m_e$ with $m_e$ being the electron mass and $E_\gamma$ being the photon energy, $\gamma$ is the Lorentz factor of the electron, and $n_{\gamma}(E_{\gamma})$ is the differential number density of ISRF photons, which is distance-dependent as it includes contributions from IR and SL.

\begin{figure}[t!]
    \centering
    \includegraphics[width=83mm]{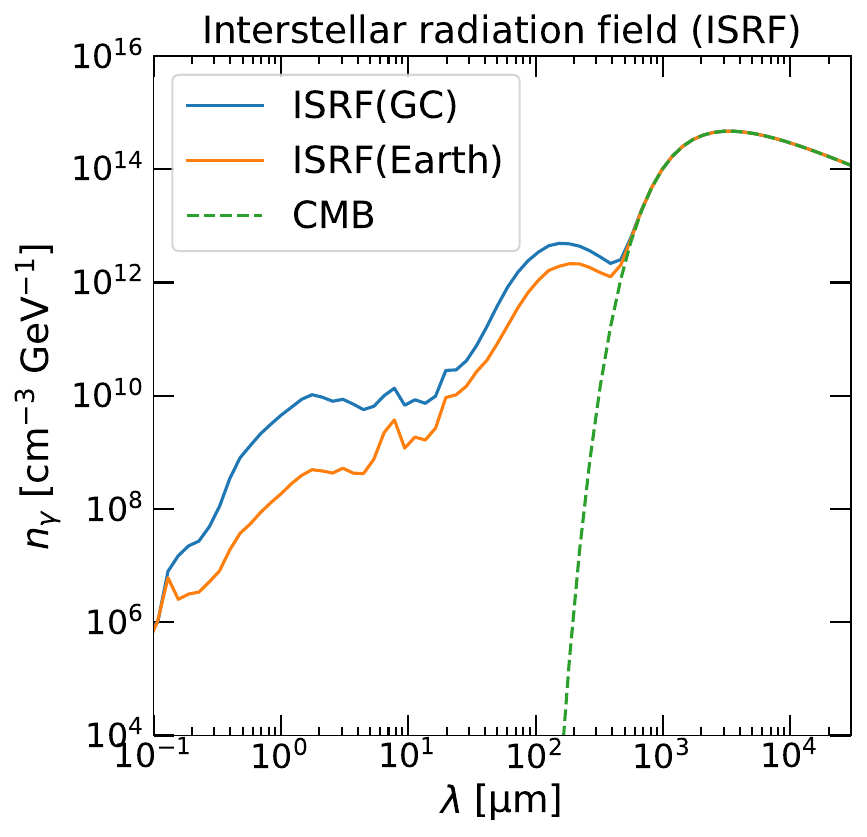}
    \caption{ISRF spectra (extracted from GALPROP~\cite{Vladimirov:2010aq}) plotted versus the photon wavelength ($\lambda$). The blue spectrum represents the ISRF at the GC, while the orange spectrum represents the ISRF at Earth. The green dashed line represents the homogeneous CMB. The three peaks from left to right correspond to SL, IR, and CMB.}
    \label{fig:.ISRF}
\end{figure}

Fig.~\ref{fig:.ISRF} shows the ISRF spectra at different positions in the Milky Way. The ISRF data is extracted from GALPROP~\cite{Vladimirov:2010aq}. The CMB spectrum remains the same in different positions as it represents a homogeneous distribution. It essentially features a blackbody spectrum with a temperature of approximately 2.73K. On the other hand, the SL and IR components are position dependent. 

The synchrotron energy loss is also distance-dependent and can be obtained as follows in cylindrical coordinates ($\rho, z$),
\begin{equation}
    b_{SYN}(\rho, z, E_e) = \frac{4 \sigma_{T} E_e^2}{3 m_e^2} \frac{B(\rho, z)^2}{2} \,.
\end{equation}
The distance dependence arises from the term $B(\rho, z)^{2}$, which represents the energy density of the galactic magnetic field. In this study, we adopt the same regular magnetic field profile as the MF1 model as that used in PPPC4DMID~\cite{Buch:2015iya},
\begin{equation}
    B(\rho,z)=B_{0} \exp\left[-\frac{\rho-R}{r_B}-\frac{|z|}{z_B}\right]\,,
\end{equation}
where R = 8.3 kpc, $r_B$ = 2 kpc and $B_0 = 4.78 \mu G$. The magnetic field profile is in cylindrical coordinates with $\rho$ and z being the radial and vertical distance from GC in cylindrical coordinates.

\begin{figure}[t!]
    \centering
    \includegraphics[width=83mm]{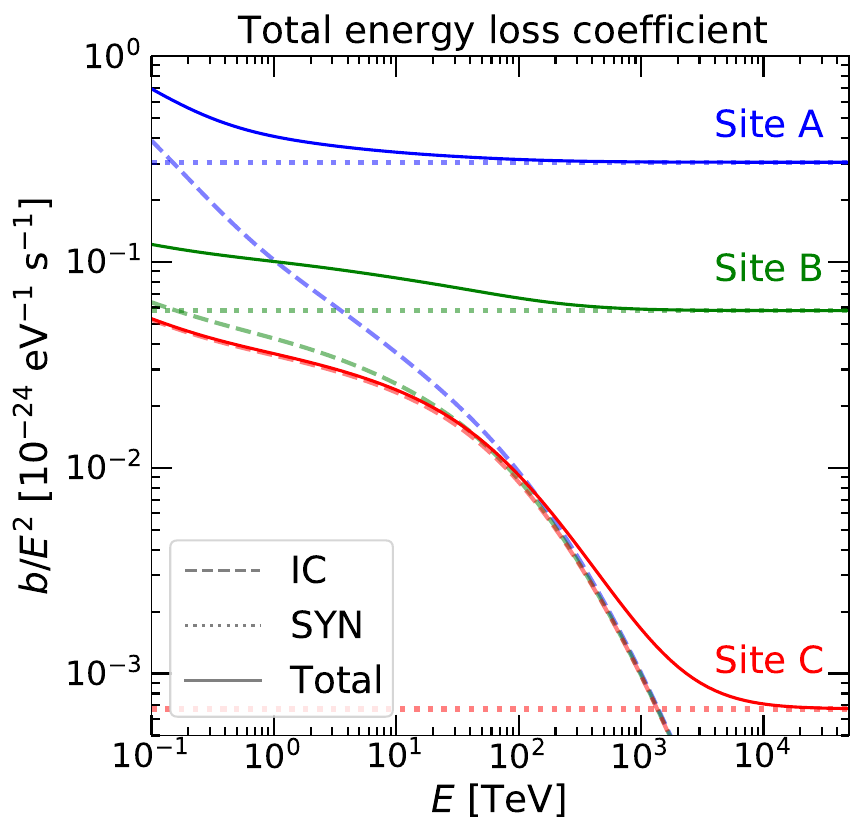}
    \caption{Total energy loss coefficient (solid line) divided by $E^{2}$ versus electron energy. The top spectrum is located at the GC (Site A); The middle is located at the Earth (Site B) and the bottom one is located at 5kpc above the GC (Site C). Individual IC components (dashed line) and synchrotron components (dotted line) are also shown.}
    \label{fig:.b_t}
\end{figure}

Fig.~\ref{fig:.b_t} shows the energy loss coefficients at several locations in the galaxy, labeled as site A (GC), site B (Earth), and site C (5\,kpc above the GC). 
In general, the IC energy loss is more important at lower energies, and the synchrotron energy loss overtakes at different energies depending on the sites. 
Moreover, it is clear to see that the total energy loss rate at the GC (site A) is the highest. This overall distance dependence arises from the galactic magnetic field profile in the synchrotron term and the ISRF profile in the IC term. Both profiles peak at the GC and decrease as the reference site moves away from it.

In addition to calculating the electron distribution, $n_e(r,E_e)$, it is also necessary to compute the IC power ($P_{IC}(E_{\gamma},E_e)$) to obtain the IC emissivity. The IC power represents the rate at which gamma rays with energy $E_{\gamma}$ are produced by a single electron with energy $E_e$ when it interacts with the ISRF. The expression of $P_{IC}$ follow the one used in PPPC4DMID~\cite{Buch:2015iya}:
\begin{multline}
    P_{IC}(E_{\gamma},E_e) = \frac{3 \sigma_{T} E_{\gamma}}{4 \gamma^2} \int^1_{1/4\gamma^2} dq \left[1-\frac{1}{4q\gamma^{2}(1-\epsilon)}\right]\\
    \frac{n_{\gamma}(E_{\gamma}^{0})}{q}\left[2q\ln q+q+1-2q^2+\frac{\epsilon^{2}(1-q)}{2(1-\epsilon)}\right]\,,
\end{multline}
where $\epsilon = E_{\gamma}/E_e$ and $E_{\gamma}^0$ is the energy for the initial photon,
\begin{equation}
    E_{\gamma}^0 = \frac{m_{e}^2 E_{\gamma}}{4qE_{e}(E_{e}-E_\gamma)}\,.
\end{equation}

With all the aforementioned components, the final galactic IC gamma-ray spectrum are obtained by performing a line-of-sight integral of the emissivity. The expression for obtaining the gamma-ray spectrum is
\begin{eqnarray}\label{eq:ICFlux}
\nonumber\frac{d^{2}\phi_{IC}}{dEd\Omega}\Big|_{decay}&=&\frac{1}{2\pi E_{\gamma}M_{DM}\tau}\int^{\infty}_{0}dx \rho_{DM}(r) \times\\     
\nonumber \int^{\frac{M_{DM}}{2}}_{E_{\gamma}}dE_{e}&&\hspace{-0.5cm}\frac{P_{IC}(E_{\gamma},E_e)}{b(E_e)}\int^{\frac{M_{DM}}{2}}_{E_{e}}dE'\frac{dN_e}{dE'}e^{-\tau_{abs}^{ISRF}}\,,\\
\nonumber \frac{d^{2}\phi_{IC}}{dEd\Omega}\Big|_{ann}&=&\frac{\langle\sigma v\rangle}{4\pi E_{\gamma}M_{DM}^2}\int^{\infty}_{0}dx \rho^{2}_{DM}(r)\times\\
\nonumber \int^{M_{DM}}_{E_{\gamma}}dE_{e}&&\hspace{-0.5cm}\frac{P_{IC}(E_{\gamma},E_e)}{b(E_e)}\int^{M_{DM}}_{E_{e}}dE'\frac{dN_e}{dE'} e^{-\tau_{abs}^{ISRF}}\,.\\
\end{eqnarray}

\subsection{Extragalactic: Prompt and Secondary Emission}
In this study, the gamma-ray energy range considered is around ${\rm TeV}$. It is worth noting that the gamma-ray signal originating from DM in the extragalactic space is also significant~\cite{Cirelli:2010xx}. At this energy range, the attenuation caused by the extragalactic background light source is not yet severe, and therefore, its contribution should be taken into account in the DM searches. The calculation for the extragalactic component should include both prompt and secondary flux, similar to the galactic component. The formalism for these calculations is similar to the Galactic component, but instead of performing a line-of-sight integral, a comoving integral should be adopted to account for the effect of redshift and the expansion of the Universe.

Considering the redshift in the extragalactic space, the modified prompt and secondary flux for decay are shown below, respectively:
\begin{eqnarray}\label{EG_decay}  
    \frac{d^{2}\phi_{prompt}^{EG}}{dEd\Omega}\Big|_{decay} &=& \frac{\Omega_{\chi}\rho_{cr}}{4\pi M_{DM}\tau}\int ^{\infty}_{0}\frac{dz}{H(z)}Att(E,z)\\
\nonumber &&   \frac{dN_\gamma}{dE}(E(1+z))\,, \\
\nonumber    \frac{d^{2}\phi_{IC}^{EG}}{dEd\Omega}\Big|_{decay} &=& \frac{\Omega_{\chi}\rho_{cr}}{2\pi E_{\gamma}M_{DM}\tau}\int^{\infty}_{0}\frac{dz Att(E,z)}{H(z)(1+z)} \\
\nonumber    \int^{M_{DM}/2}_{E_{\gamma}(1+z)}dE_{e}&&\hspace{-0.7cm}\frac{P^{CMB}_{IC}(E_{\gamma}(1+z),E_e)}{b^{CMB}_{IC}(E_e)}\int^{M_{DM}/2}_{E_{e}}dE'\frac{dN_e}{dE'}\,.
\end{eqnarray}
In the above expressions, $\Omega_{\chi}$ is chosen to be 0.27, representing the cosmological parameter for the DM density. The average cosmological DM density, $\rho_{cr}$ = $1.15 \times 10^{-6}\, {\rm GeV}\, {\rm cm^{-3}}$~\cite{ParticleDataGroup:2022pth}. Additionally, the terms $P_{IC}^{CMB}$ and $b_{IC}^{CMB}$ represent the IC emissivity and energy loss rate, respectively, which only take into account the CMB as the photon background since the full ISRF profile does not apply to the extragalactic components.

There is an attenuation factor represented by the function $Att(E,z)$, which accounts for the effect of photon-photon absorption on extragalactic background light sources. The explicit form of this attenuation factor can be found in equation (70) of the PPPC4DMID paper~\cite{Cirelli:2010xx},  where detailed information about each component's contribution can also be found. It is worth noting that the overall attenuation effect is not severe for ${\rm TeV}$ energy photons, allowing a significant portion of gamma rays to travel to Earth and be observed. For the same reason, we neglect the secondary component produced by electromagnetic cascades of the initial photon~(E.g., see Refs.~\cite{Murase:2012xs, Murase:2015gea}). 

Taking a conservative approach regarding photon attenuation, the ultraviolet (UV) background produced in the low-redshift universe when the first stars ignited is considered to be the maximum case. In this scenario, the ``maximal UV" mode is adopted, as mentioned in reference~\cite{Cirelli:2010xx}). By considering this maximal UV background, it gives the strongest gamma-ray absorption. 

Using a similar approach, the extragalactic flux for the annihilation case can be straightforwardly obtained as follows~\cite{Cirelli:2010xx}:
\begin{eqnarray}\label{EG_ann}
\nonumber    \frac{d^{2}\phi_{prompt}^{EG}}{dEd\Omega}\Big|_{ann} &=& \frac{\Omega_{\chi}\rho_{cr}^{2}\langle\sigma v \rangle}{8\pi M_{DM}^{2}}\int^{\infty}_{0} \frac{dz}{H(z)}Att(E,z)Bst(z)\\
  &&  \frac{dN_\gamma}{dE}(E(1+z))\,,\\
\nonumber    \frac{d^{2}\phi_{IC}^{EG}} {dEd\Omega^2}\Big|_{ann}&=&\frac{\Omega_{\chi}\rho_{cr}^{2}\langle\sigma v \rangle}{4\pi E_{\gamma}M_{DM}^{2}}\int^{\infty}_{0}\frac{dz Att(E,z)}{H(z)(1+z)} )Bst(z) \\
\nonumber    \int^{M_{DM}/2}_{E_{\gamma}(1+z)}dE_{e}&&\hspace{-0.7cm}\frac{P_{IC}^{CMB}(E_{\gamma}(1+z),E_e)}{b^{CMB}_{IC}(E_e)}\int^{M_{DM}/2}_{E_{e}}dE'\frac{dN_e}{dE'}\,,
\end{eqnarray}
where the extra term $Bst(z)$ represents the cosmological boost factor due to DM clumping, following the model proposed by Maccio et al.~\cite{Maccio:2008pcd}.

Finally, by summing up the galactic and extragalactic fluxes, including both the prompt and IC components, we can the total flux at specific angle is given by
\begin{eqnarray} \label{eq:total_flux}
    \frac{d^{2}\phi}{dEd\Omega}\Big|_{decay}&=&\frac{d^{2}\phi_{prompt}}{dEd\Omega}\Big|_{decay}+\frac{d^{2}\phi_{IC}}{dEd\Omega}\Big|_{decay}\\
\nonumber
    &&+\frac{d^{2}\phi_{prompt}^{EG}}{dEd\Omega}\Big|_{decay}+\frac{d^{2}\phi_{IC}^{EG}}{dEd\Omega}\Big|_{decay}\,, \\
\nonumber
    \frac{d^{2}\phi}{dEd\Omega}\Big|_{ann}&=&\frac{d^{2}\phi_{prompt}}{dEd\Omega}\Big|_{ann}+\frac{d^{2}\phi_{IC}}{dEd\Omega}\Big|_{ann}\\
\nonumber
    &&+\frac{d^{2}\phi_{prompt}^{EG}}{dEd\Omega}\Big|_{ann}+\frac{d^{2}\phi_{IC}^{EG}}{dEd\Omega}\Big|_{ann}\,.
\end{eqnarray}

\begin{figure}[t!]
    \centering
    \includegraphics[width=83mm]{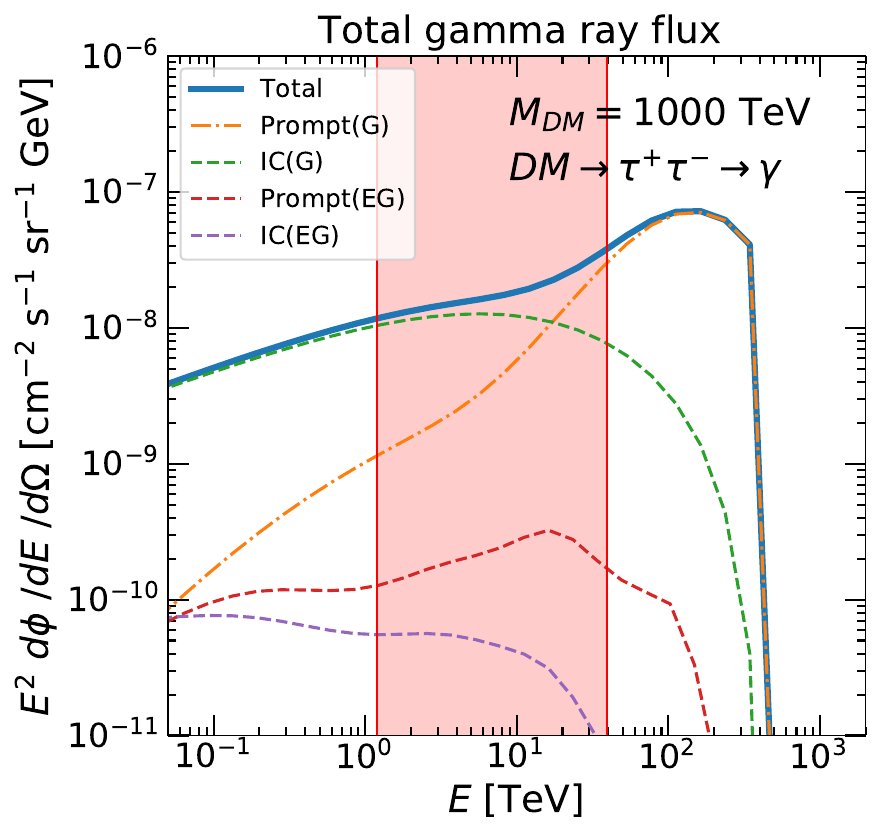}
    \caption{The total gamma-ray flux from DM decay is shown in blue along with the individual contributions of the galactic prompt flux in orange, galactic secondary flux in green, extragalactic prompt flux in red, and extragalactic secondary flux in purple. The observation angle (DEC, RA) is set to ($2.5^\circ, 240^\circ$). The specific channel chosen is $DM \rightarrow \tau^+\tau^-$, with a DM mass of $1000\,{\rm TeV}$ and a lifetime of $1\times10^{27}\,{\rm s}$. The red shaded region represents the energy range of interest in this study, set by the energy range of the HAWC flux upper limits~\cite{HAWC:2017udy}.}
    \label{fig:totcomp_tau}
\end{figure}

Fig.~\ref{fig:totcomp_tau} shows the plot representing the total gamma-ray flux resulting from DM decay via the tau channel ($DM \rightarrow \tau^+\tau^-$), using the methodology discussed above. The contributions from both galactic and extragalactic components are depicted separately.

In this specific case, the secondary fluxes are found to be higher than the prompt fluxes for photon energies $\lesssim 10^4\,{\rm GeV}$. As a result, the dominant source of gamma rays in the lower energy range is the IC flux, both from the galactic and extragalactic components. 
In particular, when the DM mass is above the energy range where the data is available, the IC component can even dominate over the prompt component, and thus is important for setting the sensitivity.

\begin{figure}[t!]
    \centering
    \includegraphics[width=83mm]{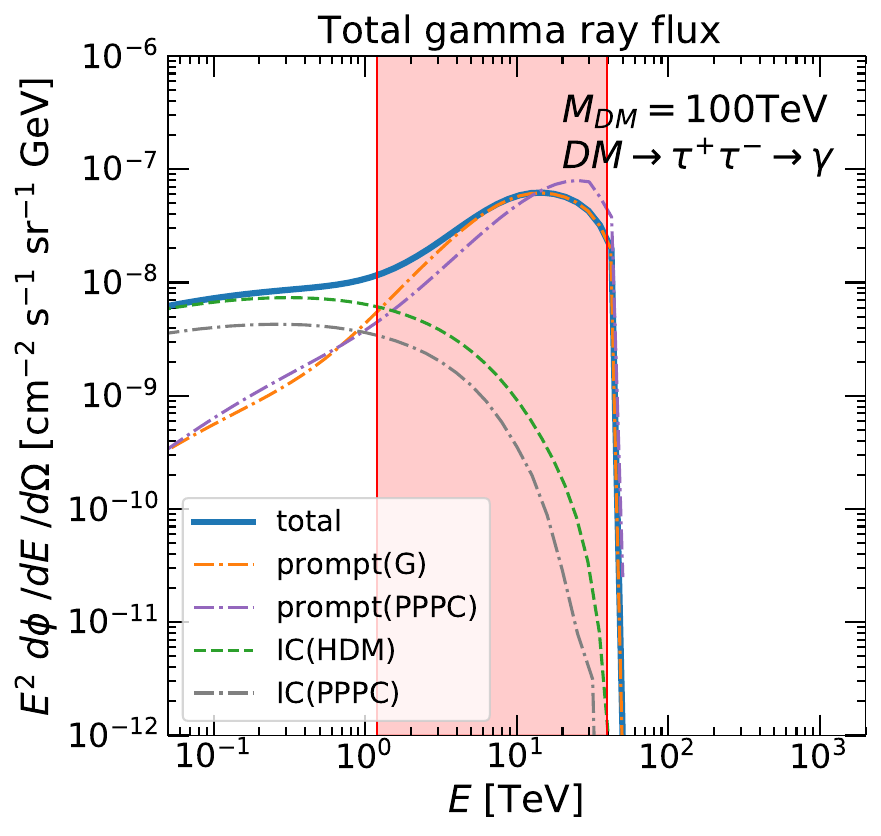}
    \includegraphics[width=83mm]{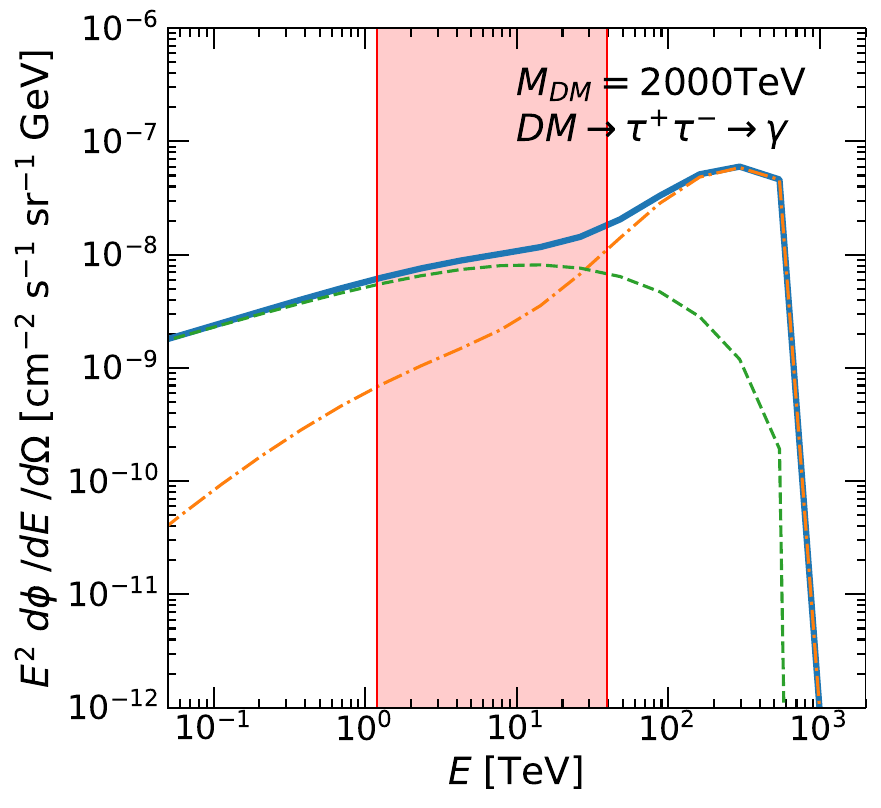}
    \caption{Same as Fig.~\ref{fig:totcomp_tau}, but for 100\,TeV~(top panel) and 2000\,TeV~(bottom panel) DM masses, and only showing the Galactic components. In the top panel, we compare with our Galactic prompt and IC results that use HDM versus that from PPPC4DMID~\cite{Cirelli:2010xx}, which are shown in purple (prompt) and grey (IC), respectively. The parameters are chosen to be the same as Fig.~\ref{fig:totcomp_tau} except DM mass values.}
    \label{fig:.totcomp}
\end{figure}

Fig.~\ref{fig:.totcomp} shows the galactic component (prompt + IC) of the gamma-ray flux for $100\,{\rm TeV}$~(top panel) and $2000\,{\rm TeV}$~(bottom panel) DM masses. 
In the top panel, we also show the IC flux from the PPPC4DMID~\cite{Cirelli:2010xx}. 
Our prompt flux results agree well with that from PPPC4DMID, with minor difference coming from the different source of the single particle spectrum. 
For the secondary IC flux, an additional source of difference is the treatment of electron and positron diffusion. 
Here we compared our work with the "MAX" model defined in \cite{Buch:2015iya} which represents the scenario that maximizes the final flux after incorporating diffusion. 

In the bottom panel of Fig.~\ref{fig:.totcomp}, we show that when the DM mass is increased, the entire spectrum generally shift towards higher energies. 
However, in this work, we are only interested in the specific energy range in the red shaded region, limited by the HAWC observation. It is then clear that the IC contribution is important.    

\section{Method}\label{sec:Method}
\subsection{The High Altitude Water Cherenkov Observatory (HAWC)}
HAWC is a ground-based gamma-ray observatory. It consists of an array of 300 water-Cherenkov detectors (WCDs). Each WCD is a steel tank that contains purified liquid water and is equipped with four photomultiplier tubes (PMTs) attached at the bottom for the detection of Cherenkov radiation. Cherenkov radiation is emitted when high-energy particles from an air shower in the atmosphere enter the water. By applying gamma/hadron cuts, HAWC is able to distinguish between air shower events triggered by high-energy gamma rays or hadronic cosmic rays. The direction of the gamma ray or cosmic ray responsible for the air shower event is then reconstructed. The detailed procedure for this reconstruction can be found in Ref.~\cite{HAWC:2014ics}. Regarding the energy of the observed gamma-ray, it is determined by the fraction of functioning PMTs triggered in an air shower event.

\subsection{Overview of HAWC 2018 flux limits}
\begin{figure}[t!]
    \centering
    \includegraphics[width=70mm]{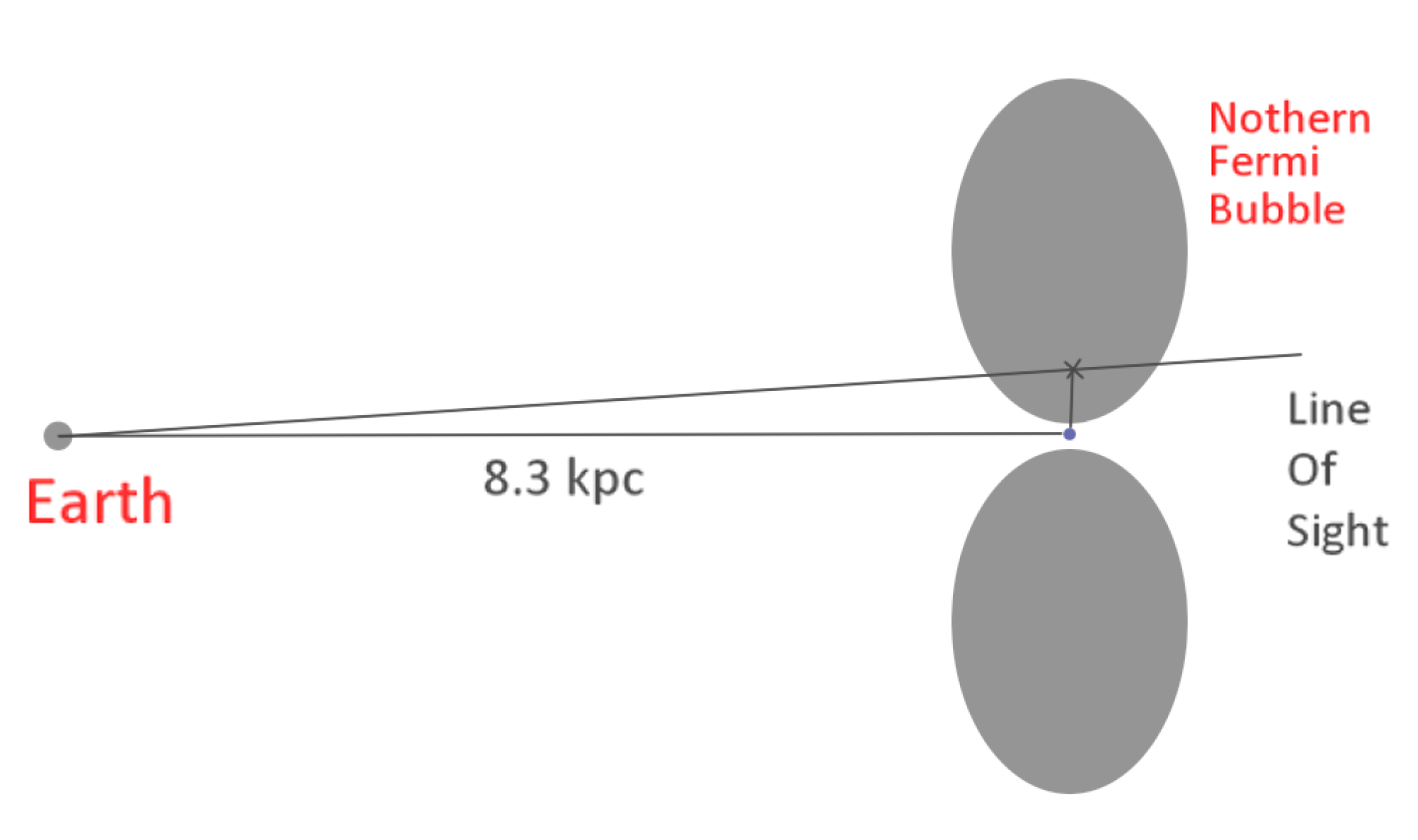}
    \caption{Configuration of the line of sight path looking into the Northern Fermi Bubble from the Earth.}
    \label{fig:.config}
\end{figure}

We consider the $95\%$ confidence level (CL) upper limits reported by HAWC observatory between November 27th, 2014, to February 11th, 2016, as reported in their 2017 and 2018 papers~\cite{Abeysekara:2017wzt,HAWC:2017udy}. The chosen sky area in this paper is the Northern Fermi bubble as it is in closer proximity to the GC. In practice, the search region of the Northern Fermi Bubble is divided into seven declination bands, each spanning $5^\circ$ each~(see Table~\ref{table:1}). The overall configuration can be referred to Fig. \ref{fig:.config}. The expected flux in each declination band is assumed to be pointing at the midpoint of that band in this work, with the right ascension set at $240^{\circ}$.

\begin{table}
\begin{tabular}{ |p{1cm}|p{2.2cm}|  }
\hline
\multicolumn{2}{|c|}{Declination bins} \\
\hline
Bin& (DEC, RA) \\
\hline
1 & ($6.5^\circ$,$240^\circ$) \\
2 & ($2.5^\circ$,$240^\circ$) \\
3 & ($-2.5^\circ$,$240^\circ$) \\
4 & ($-7.5^\circ$,$240^\circ$) \\
5 & ($-12.5^\circ$,$240^\circ$) \\
6 & ($-17.5^\circ$,$240^\circ$) \\
7 & ($-22.5^\circ$,$240^\circ$) \\
\hline
\end{tabular}
\caption{The equatorial coordinate of each declination bin of the Northern Fermi bubble being observed, corresponding to the 7 angular regions reported in HAWC 2018~\cite{HAWC:2017udy}.}
\label{table:1}
\end{table}

\subsection{Reinterpreting the HAWC limits}

The HAWC flux upper limits~\cite{HAWC:2017udy} that we consider is listed in Table~\ref{table:2}. 
A total of 21 data points are available, which can be employed to constrain the DM lifetime when combined with our calculated flux. 

The parameters of interests are the decay rates $\Gamma = 1/\tau$ or the annihilation cross section $\langle\sigma v\rangle$.  We denote the flux upper limits by HAWC in the $i$-th energy bin and the $j$-th angular bin as $F^{lim}_{ij}$. The expected DM flux is then $F^{model}_{ij}(\Gamma~ {\rm or}~ \langle\sigma v\rangle)$, which either depends on the decay rate or the annihilation cross section. 
$F^{model}_{ij}$ is obtained from $d^2\phi/dEd\Omega$ in Eq.~\ref{eq:total_flux} after averaging over the corresponding energy and angular bins.

Following the same approach as the Dwarf analysis by HAWC~\cite{HAWC:2017mfa} (and see Appendix~\ref{Appendix} for details), the test statistics (TS) for each bin is given by
\begin{equation}\label{eq:ts_bin}
    TS=2.71\left(\frac{ F^{model}_{ij}}{F^{lim}_{ij}}\right)^{2}\,,
\end{equation}
and the total TS for combining the 21 angular and energy bins is 
\begin{equation}\label{eq:TStot}
    TS^{tot}(\Gamma~ {/}~ \langle\sigma v\rangle)=\sum_{i,j} 2.71\left(\frac{F^{model}_{ij}}{F^{lim}_{ij}}\right)^{2}\,.
\end{equation}
The 95\% confidence level (CL) limit on the decay rate or the annihilation cross section can be obtained by solving for $TS^{tot} = 2.71$. 

\section{Result and Discussion}\label{sec:result}
\begin{table*}
  \centering
  
  \begin{tabular}{ |p{2.2cm}|p{2.2cm}|p{2.2cm}|p{2.2cm}|p{2.2cm}|p{2.2cm}|p{2.2cm}|p{2.2cm}|  }
    \hline
    \multicolumn{8}{|c|}{Fluxes and Upper limits per Declination Band ($E^2d^2F/dEd\Omega$) ($\times 10^{-7} \rm  GeVcm^{-2}s^{-1}sr^{-1}$)} \\
    \hline
    Energy bin& $5^{\circ} - 8^{\circ}$ & $0^{\circ} - 5^{\circ}$ & $-5^{\circ} - 0^{\circ}$ & $-10^{\circ} - -5^{\circ}$ & $-15^{\circ} - -10^{\circ}$ & $-20^{\circ} - -15^{\circ}$ & $-25^{\circ} - -20^{\circ}$ \\
    \hline
    $[1.2-3.9]$   & $2.8\pm6.3$ & $-4.3\pm3.4$ & $5.9\pm3.8$ & $-0.8\pm5.0$ & $-4.5\pm8.0$ & $-31.0\pm16.0$ & $-57.0\pm42.0$\\
    $[3.9-12.4]$  & $-0.9\pm2.3$ & $-1.1\pm1.2$& $2.8\pm1.2$ & $-0.4\pm1.5$ & $-0.9\pm2.0$ & $-5.9\pm3.1$ & $-12.0\pm6.6$\\
    $[12.4-39.1]$ & $-1.2\pm1.3$ & $-0.1\pm0.6$ & $-0.3\pm0.6$ & $0.1\pm0.7$ & $0.0\pm0.8$ & $-1.6\pm1.1$ & $-2.6\pm2.0$\\
    \hline
  \end{tabular}
  \caption{The gamma-ray energy bin and flux upper limits provided by the HAWC collaboration~\cite{HAWC:2017udy}.}
  \label{table:2}
\end{table*}

\begin{figure}[h]
\minipage{0.33\textwidth}
    \includegraphics[width=62mm]{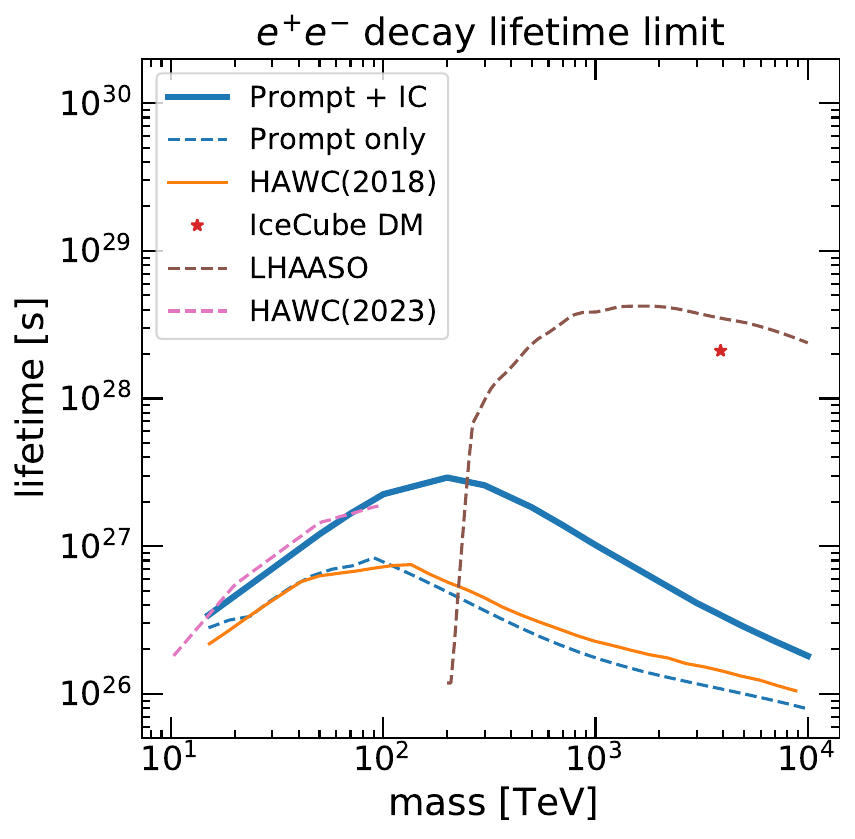}
\endminipage\hfill 
\minipage{0.33\textwidth}
    \includegraphics[width=62mm]{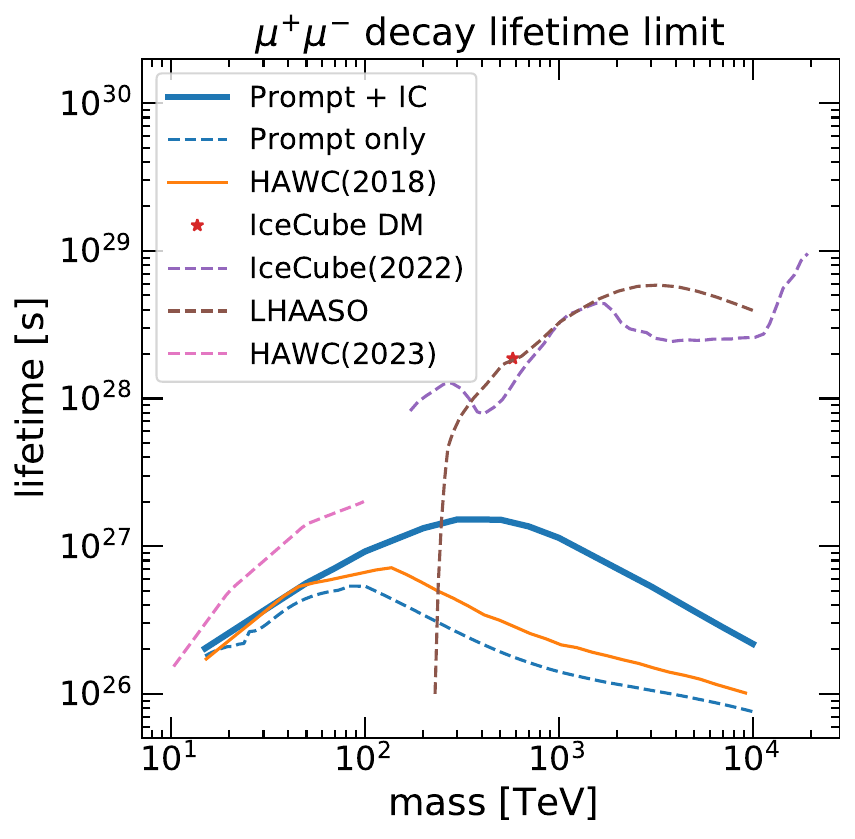}
\endminipage\hfill 
\minipage{0.33\textwidth}
    \includegraphics[width=62mm]{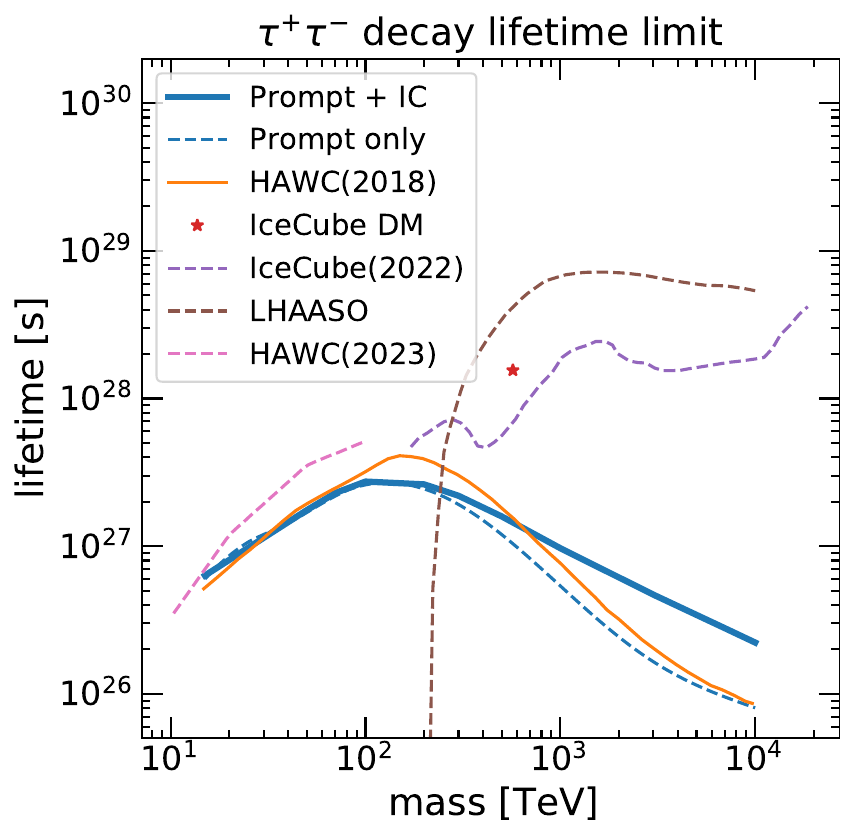}
\endminipage\hfill 
    \caption{The DM decay lifetime limits for leptonic channels. Our results are shown in blue solid lines and our prompt-only results in blue dashed lines. For comparison, we also show the previous results from HAWC in 2018 (Orange)~\cite{HAWC:2017udy}, HAWC 2023 (Pink)~\cite{HAWC:2023owv}, the IceCube limits (Purple)~\cite{IceCube:2022clp}, recent LHAASO limits (Brown)~\cite{LHAASO:2022yxw}, and the best fit favored by IceCube high-energy neutrino flux (Red stars)~\cite{Chianese:2019kyl}.}
    \label{fig:lepton_ice}

\minipage{0.33\textwidth}
    \includegraphics[width=62mm]{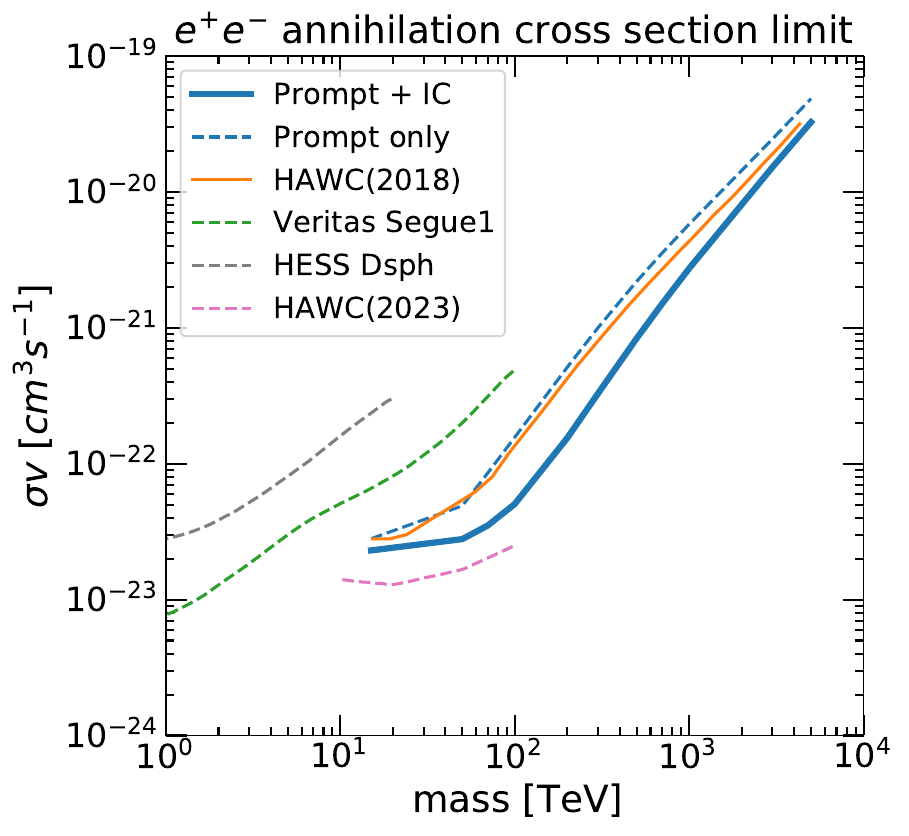}
\endminipage\hfill 
\minipage{0.33\textwidth}
    \includegraphics[width=62mm]{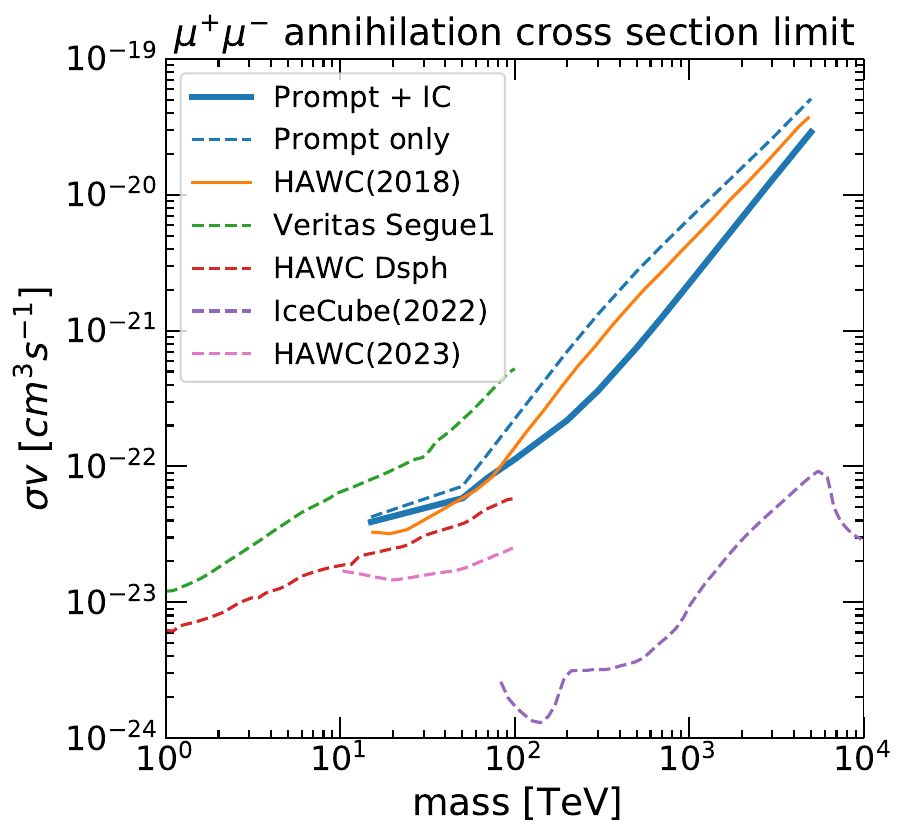}
\endminipage\hfill 
\minipage{0.33\textwidth}
    \includegraphics[width=62mm]{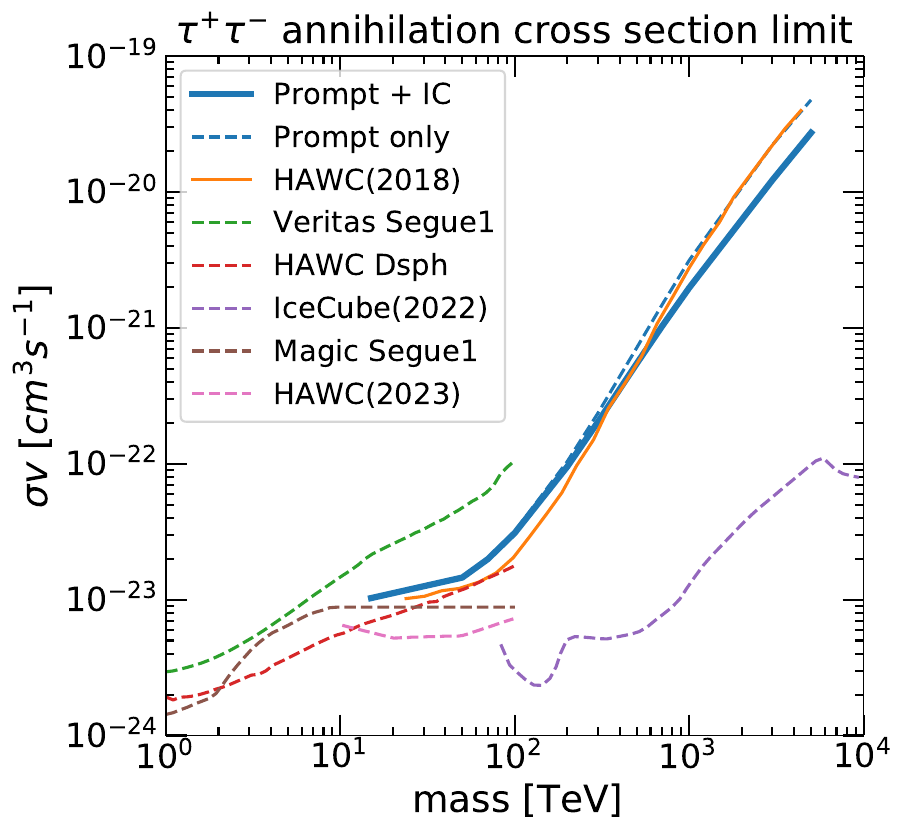}
\endminipage\hfill 
    \caption{Same as Fig.~\ref{fig:lepton_ice}, but for DM annihilation cross sections. We also show the results from HAWC in 2018 (Orange)~\cite{HAWC:2017udy}, HAWC 2023 (Pink)~\cite{HAWC:2023owv}, Segue 1 dwarf galaxy limits from Veritas (Green)~\cite{VERITAS:2017tif}, MAGIC dwarf analysis (Brown)~\cite{MAGIC:2016xys}, HESS dwarf analysis (Grey)~\cite{HESS:2014zqa}, HAWC (Red)~\cite{HAWC:2017mfa}, and the IceCube limits (Purple)~\cite{IceCube:2023ies}.}
    \label{fig:lepton_ann}
\end{figure}

Fig.~\ref{fig:lepton_ice} and Fig.~\ref{fig:lepton_ann} show the limits for DM decay lifetime and annihilation cross section for the leptonic channels ($e, \mu, \tau$).  The limits obtained including both prompt and IC are shown in thick blue lines, while the prompt-only limits are shown with thin blue dashed lines. Compared to the original HAWC limits from 2018~\cite{HAWC:2017udy}, our prompt-only results generally agree well with the HAWC results.  The minor differences are caused by the differences from the final state spectrum~($dN/dE$) used between these two works. 

As can be seen from Fig.~\ref{fig:lepton_ice} and Fig.~\ref{fig:lepton_ann}, the inclusion of the IC component improves the constraints.  For channels that are less rich in prompt gamma rays (e.g., $e$ and $\mu$), the improvements can be substantial, by almost one order of magnitude in some DM mass values. There are also some differences between the improvements form decay versus annihilation, which is due to the difference in the spatial dependence of the IC contribution. We also note that part of the improvements comes from the fact that our limits are derived from flux upper limits in a fixed energy range.  As can be seen from Fig.~\ref{fig:.totcomp}, for DM masses above the energy range considered, the relative importance of the secondary and EG components increases. The secondary IC contribution therefore allows the detectors to be sensitive to a wider range of DM masses. 

We also compare with results from other experiments.  For DM decays, we compare with the recent limits by LHAASO-KM2A~\cite{LHAASO:2022yxw} as well as neutrino limits by IceCube~\cite{IceCube:2022clp}, whenever the limits are available for the specific channel. In general, the limits from these experiments are stronger at higher DM masses.  But HAWC, with a lower energy threshold, sets the strongest limits at lower DM masses~(roughly below 100\,TeV).

For DM annihilation, we compare with existing results from VERITAS, MAGIC, HESS, and HAWC dwarf galaxy analysis~\cite{VERITAS:2017tif, MAGIC:2016xys, HESS:2014zqa, HAWC:2017mfa}.  Generally the HAWC Galactic halo limits are competitive around 10\,TeV, and can extend to higher DM masses. But for neutrino rich channels, such as $\mu$ and $\tau$ channels, IceCube~\cite{IceCube:2023ies} provides stronger constraints. 

In addition to the results from HAWC 2018, we also compare with recent results by HAWC 2023~\cite{HAWC:2023owv}, which performed a more optimised DM search with the Galactic halo observation, but the secondary IC contribution is also not considered.  We note that the flux upper limits are not provided in HAWC 2023, thus we cannot perform the analysis in this work with the updated result from HAWC 2023. Generally, the HAWC 2023 results improves by a factor of a few compared to HAWC 2018. 
As demonstrated in this work, we expect a similarly degree of improvements in the limits if the IC and extragalactic components are included in this analysis.  

\begin{figure}[h]
\minipage{0.33\textwidth}
   \includegraphics[width=62mm]{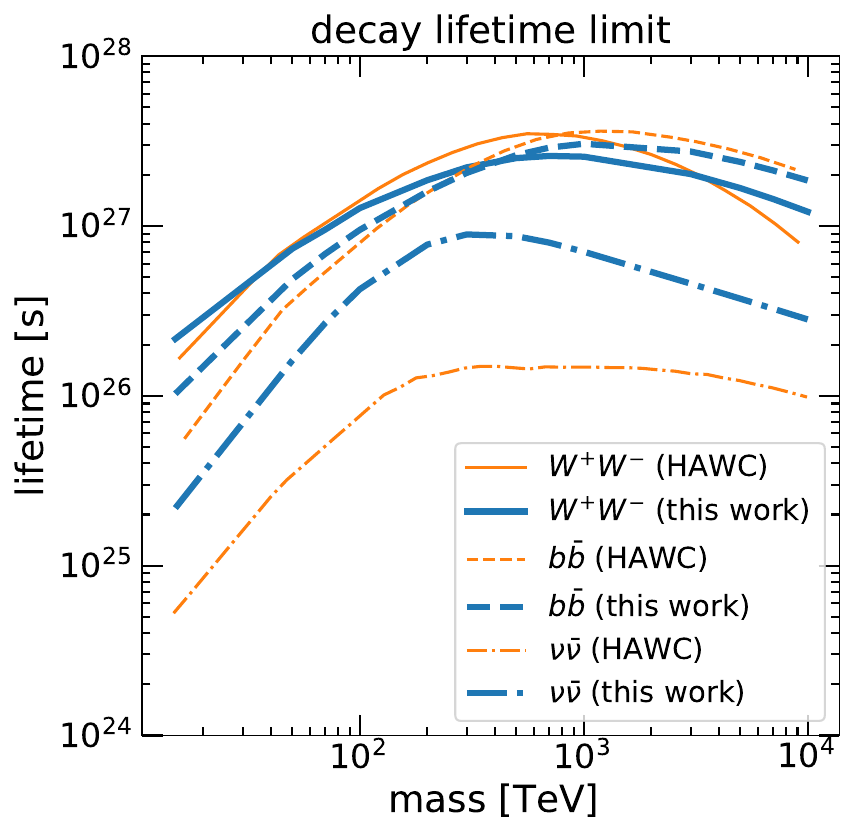}
\endminipage\hspace{0.5cm}
\minipage{0.33\textwidth}
    \includegraphics[width=62mm]{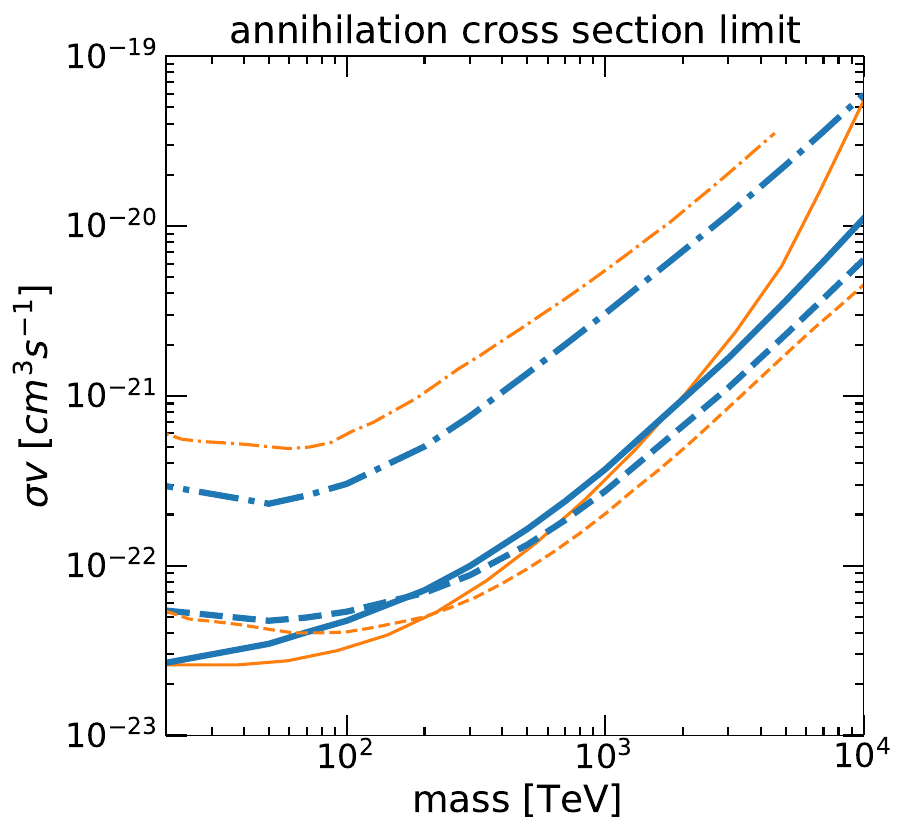}
\endminipage\hspace{0.5cm}
    \caption{The DM decay lifetime (left) and annihilation cross section limits (right) for $W^+W^-$, $b\bar{b}$, and $\nu\bar{\nu}$ channels. The 2018 HAWC limits~\cite{HAWC:2017udy} are plotted in dashed lines for comparison.}
    \label{fig:.bbww}
\end{figure}

Fig.~\ref{fig:.bbww} shows our results for hadronic~($W$ and $b$) and neutrino channels. We only compare with the results from HAWC 2018 here to highlight the impact of including secondary IC contributions. As expected, for these hadronic channels that are rich in prompt gamma rays, the secondary IC contribution does not significantly improve the results. However, the neutrino channels, the improvement is substantial. 

\newpage   

\onecolumngrid

\twocolumngrid

\section{Conclusions and Outlook}\label{sec:Conclusions}
In this work, we consider the contribution of secondary IC emission and extragalactic prompt and IC contribution in Galactic halo DM searches. We include these contributions and re-interpret the flux upper limits provided in HAWC 2018~\cite{HAWC:2017udy} to obtain new limits on DM decay lifetime and annihilation cross section.  Depending on channels, we see improvements on the limits at the level of a factor of few to one order of magnitude. This highlights the importance of considering the secondary contributions, particular for channels that have weak prompt gamma-ray emission. 

We demonstrate the importance of the IC component in this work using a simplified IC formalism that assumes energy loss dominates over diffusion. This approximation is expected to be more accurate at high energy due to the energy dependence of the energy loss process.  For future works, the diffusion effects could be included and systematic studies of the IC components against various propagation scenario should be investigated~(e.g., similar to Ref.~\cite{Buch:2015iya}).  Furthermore, it was pointed out recently that stochastic effects may be important in DM studies~\cite{John:2023ulx}. 

In the near future, LHAASO-WCDA~\cite{LHAASO:2019qtb} should accumulate more data than HAWC given its larger effective area. Furthermore, the proposed SWGO detector~\cite{swgo:2019ahw} should have much better DM sensitivity given that it can observe southern hemisphere that contains the GC. Both of these should have better DM sensitivities than HAWC, and will be important tools for DM searches with high-energy astrophysical gamma rays.

% ====== acknowledgments ====== %
\bigskip
\section*{Acknowledgments}
The works of DMHL and KCYN are supported by Croucher foundation, RGC grants (24302721, 14305822, 14308023), and NSFC/GRC grant (N\_CUHK456/22).

\bibliography{bib.bib}
\newpage
\appendix

\section{Test Statistics from Flux limits}\label{Appendix}
Following the same idea from Ref.~\cite{HAWC:2017mfa}, assuming that the TS value for a particular bin satisfies the chi square formula,
\begin{equation}
    TS=\left[\frac{(F^{model}+F^{BG})-F^{D}}{\sigma}\right]^{2} \,,
\end{equation}
where $F^{model}$ is the model flux contribution, $F^{BG}$ is the modeled background flux, $F^{D}$ is the measured flux by the detector, and $\sigma$ is the uncertainty of this bin. Assuming that the background model flux can reasonably describe the observed flux (i.e., $F^{D} \approx F^{BG}$), the 95\% flux upper limit, $F^{lim}$, is given by
\begin{equation}\label{eq:2.71TS}
    2.71 = \left(\frac{F^{lim}}{\sigma}\right)^{2}\,.
\end{equation}
Eliminating $\sigma$ from the above two equations and using $F^{D} \approx F^{BG}$, Eq.~\ref{eq:ts_bin} can be obtained.

\end{document}